\documentclass{aastex}

\slugcomment{E00-5682, buta.0810
to appear in the Astrophysical Journal}

\def\arcs{$^{\prime\prime}$}

\shorttitle{Buta and Block}
\shortauthors{Dust-Penetrated Bar Classification}

\begin{document}


\title{A Dust-Penetrated Classification Scheme For Bars\\
 As Inferred From Their Gravitational Force Fields}


\author{R. Buta\altaffilmark{1} and D. L. Block}
\affil{Department of Computational and Applied Mathematics, University of the
Witwatersrand, Johannesburg, South Africa}

\altaffiltext{1}{Permanent Address: Department of Physics and Astronomy,
University of Alabama, Box 870324, Tuscaloosa, Alabama 35487, USA}

\begin{abstract}

The division of galaxies into ``barred'' (SB) and ``normal'' (S) spirals is a
fundamental aspect of the Hubble galaxy classification system. This
``tuning fork'' view was revised by de Vaucouleurs, whose
classification volume recognized apparent ``bar strength'' (SA, SAB, SB)
as a continuous property of galaxies called the ``family''.
However, the SA, SAB, and
SB families are purely visual judgments that can have little bearing
on the actual bar strength in a given galaxy. Until very recently,
published bar judgments were based exclusively on blue light images,
where internal extinction or star formation can either mask a bar
completely or give the false impression of a bar in a nonbarred galaxy.

Near-infrared camera arrays, which principally trace the old stellar
population in both normal and barred  galaxies, now facilitate a quantification of bar
strength in terms of their gravitational potentials and force
fields.  In this paper, we show that the maximum
value, $Q_b$, of the ratio of the tangential force to the mean
axisymmetric radial force in a barred disk galaxy is a quantitative
measure of the strength of a bar. $Q_b$ does not measure bar
ellipticity or bar shape, but rather depends on the actual forcing due
to the bar embedded in its disk.   We show that a wide
range of true bar strengths characterizes the category ``SB'', while
de Vaucouleurs category ``SAB'' corresponds to a much narrower range of
bar strengths. We present $Q_b$ values for 36 galaxies, and we incorporate
our bar classes into a dust-penetrated classification system for
spiral galaxies.

\end{abstract}


\section{Introduction}

The presence of a bar in a disk galaxy implies a non-axisymmetric gravitational
field.  The high frequency of occurrence of bars (over 65 percent,
de Vaucouleurs 1963; Eskridge et al. 2000) and the fact that bars principally
consist of an old stellar population \citep{dev55,dvv59,fre89,elm85}
implies that bars are fundamental
components in the distribution of mass in galaxies.

Bars are believed to be the ``engines'' driving a wide variety of
secular evolution processes in galaxy dynamics \citep{pfe96}. Bar
driven secular evolution appears to make significant changes in galaxy
structure over a Hubble time. The main components found in barred
galaxies include bulges, disks, lenses, and inner and outer rings -- and
of course the bar itself \citep{san61,dev59,kor79,sel93,but95,bco96}.

The absence or presence of a bar led \citet{hub26} to develop two
separate prongs to his classification tuning fork: the sequence of normal
(unbarred) spirals Sa, Sb, and Sc,
paralleled by a sequence of barred spirals SBa, SBb and SBc.
\citet{dev59} recognized that galaxies such as NGC 5236 (M83)
showed a bar morphology intermediate between that of a normal and a
barred spiral. He introduced the notation SA for unbarred spirals so
that he could use the combined notation SAB for transitional cases
like M83.

In this paper, we recognize a full continuum of ``bar strengths'', as does
de Vaucouleurs, but we propose a numerical quantification of bar strength based
on the gravitational forcing of the bar itself, not on visual appearance.
From a sample of 36 galaxies, we recognize seven bar strength classes:
bar class 0 galaxies, which are normal spiral galaxies without any bar; bar
class 1 and 2 galaxies, which show ovals and weak bars (which de Vaucouleurs
would have classified in his SAB class); and bar class 2-6 galaxies,
which encompass all galaxies classified as SB by Hubble and de Vaucouleurs.

Our numerical quantification of bar strength is derived from the
non-axisymmetric force field of a galaxy inferred from the near-infrared light
distribution. Near-infrared $H$-band (effective wavelength 1.6$\mu m$)
and $K$-band (effective wavelength 2.2 $\mu m$) images
beautifully reveal the old stellar
population or `backbone' of spiral galaxies \citep{fro96,blo94,blo99}.
The near-infrared light
comes principally from old giant and supergiant stars \citep{fro96}. The
extinction at $H$ and $K$ is only 0.1-0.2 times that in visual light,
so that dust
has only a minimal effect on the inferred potentials. A rich duality of
spiral structure has been found from studies of optical and
near-infrared images; a spiral galaxy may present two completely different
morphologies when examined optically and in the near-infrared
\citep{elm99,blo99}.

In the optical, dust often
hides bars, as in the Milky Way. Seventy percent of spirals classified
in the {\it Carnegie Atlas of Galaxies} \citep{san94}, based on Hubble
bins, are classified as unbarred. This fraction drops to
27\% when these galaxies are imaged in the near-infrared \citep{esk99}.
The remaining 43 percent show ovals or bars in the
dust-penetrated regime. This high percentage of bars in the near-infrared
agrees with the findings of \citet{sei98},
who found that 90\% of a sample of 45 galaxies showed
some evidence of a bar in the $K$-band.

\section{Bar Strength as a Measurable Parameter of Galaxies}

A variety of quantitative parameters has been suggested or could
be interpreted to represent a
measure of the ``strength'' of a bar. The bar-interbar contrast,
developed by \citet{elm85}, can distinguish strong bars from weak bars, but
may connect only indirectly to an actual bar strength. Sometimes the
maximum bar-interbar contrast occurs inside the radius of the bar,
but in some cases it may occur outside the ends of the bar,
as in NGC 1433 \citep{but86}. Also, \citet{sei98} note that
bar-interbar contrasts may be weakened by resolution and seeing effects.

\citet{elm85} also used
Fourier intensity amplitudes to derive relative bar luminosities
in terms of $m$ = 2 and 4 components. Relative to the total luminosity
of the disk within the standard isophote having $\mu_B$ = 25.0 mag
arcsec$^{-2}$, the bar luminosity fraction was found to range from less
than 1\% to more than 20\%. \citet{oht90} derive similar
parameters for six barred galaxies, including the $m$ = 6 term.

The most popular bar strength parameter, because
of its simplicity, is the deprojected bar ellipticity, $\epsilon_b$,
developed by \citet{mar95} and suggested by analytical models
\citep{ath92} to be
a readily accessible measure of the strengths of bars that does not
depend on spectroscopic observations, surface photometry, or
mass-to-light ratio assumptions. \citet{mar95} derived $\epsilon_b$
for more than 100 spiral galaxies by visual inspection of published
optical photographs, and noted that the slope of chemical abundance
gradients \citep{mar94} as well as the presence of nuclear
star formation \citep{mar97} depends on $\epsilon_b$.
This parameter (or its equivalent, $(b/a)_{bar}$)
has also been used in a number of other recent papers
\citep{roz98,agu99,cha99,abr99,shl00}. \citet{abr99}
describe an algorithm that automatically derives $(b/a)_{bar}$
from moments of the galaxy image taken at various cuts
relative to the maximum flux level. This approach is useful
for high redshift morphological studies.
\citet{abr00} describe a refinement on bar axis
ratio using a parameter $f_{bar}$, defined as ``the minimum
fraction of the bar's stars that one would have to rearrange in order
to transform the structure into an axisymmetric distribution.''
\citet{mar95} notes that $\epsilon_b$ is not a complete description of
bar strength, but merely the most accessible one.

\citet{sei98} developed another quantitative approach to bar
strength that utilized near-infrared surface photometry. In their
method, the bar is defined to be the light remaining after disk
and bulge components are subtracted. This light is converted into a
parameter known as ``equivalent angle'' (EA) which is defined to be
``the angle subtended at the centre of the galaxy by a sector of the
underlying disk and bulge that emits as much light as the bar component,
within the same radial limits.'' An advantage of this method is that
it accounts for both conventional bars as well as ovals. However,
the method relies on full bulge/disk
photometric decompositions which can be very difficult for strongly-barred
galaxies. \citet{roz98} derived another flux
parameter, $\sigma_b$, representing the ratio of the flux inside the bar
to that outside the bar area. They argue that this parameter and $\epsilon_b$
indicate that stronger bars are accompanied by a lower degree of symmetry of
star formation in the spiral arms.

In each of these methods, the bar itself has to be defined,
e.g., where it appears to start, where it appears to end, or
where the maximum ellipticity is achieved.
Our approach here is different, and is based instead
on the torques induced by the
rigidly rotating bar,  without first having to accurately
define and isolate it relative to the other components in a galaxy.

\section{Bar Strength as a Force Ratio Rather than an Axis Ratio}

The most elegant way of measuring the torques of bars embedded
in disks is actually an old idea. Given the gravitational
potential $\Phi(R,\theta)$ in the disk plane,
\citet{com81} proposed defining the bar strength at radius $R$ as

\begin{equation}
Q_T(R) = {F_T^{max}(R)\over <F_{R}(R)>}
\end{equation}

\noindent
where $F_T^{max}(R)$ = $({\partial\Phi(R,\theta)/\partial\theta})_{max}$
represents the maximum amplitude of the tangential force
at radius $R$ and $<F_{R}(R)> =
R (d\Phi_0/dR)$ is the mean axisymmetric radial
force at the same radius, derived from the $m$ = 0 component of the
gravitational potential. Although $Q_T$ depends on radius, the maximum
value of $Q_T$ can provide a single measure of bar strength for
a whole galaxy, if the gravitational potential is known. With
the advent of near-infrared arrays, and the availability of many
high quality $K$ and $H$-band images of galaxies covering a range
of apparent bar morphologies, it has become possible for the first
time to use the idea in equation 1 in a practical way to
directly measure the strengths of bars from their force fields
for a large number of galaxies. This provides us with an opportunity
to develop a
consistent and robust scheme of dust-penetrated classification of bars
embedded within disks. We outline our application of equation 1 here in a
preliminary way using a small sample of near-infrared images, and discuss
its advantages and disadvantages over the methods just described.
Our subset of near-infrared images comes mainly from a much larger sample
used to develop a near-infrared classification scheme of spiral
galaxies, based on a wide variety of telescopes and detectors.
Except for Maffei 2 and M101, full details
of the observing procedures, integration times, filters and image
scales have been described and published in \citet{blo94,blo99}
and references therein. The images of Maffei 2 and M101 that we have used
are from the Two Micron All Sky Survey (2MASS; Jarrett 2000), and have an image
scale of 1\arcs\ pix$^{-1}$.

\section{Gravitational Potentials of Galaxies from Near-Infrared Images}

\citet{qui96} reviews the principal issues in deriving gravitational
potentials from near-infrared images.
The principal assumption is that the light traces the mass,
i.e., the mass-to-light ratio is constant across much of the disk.
Then the potential in two dimensions can be derived as the convolution of
the mass density with the function $1/R$ using fast Fourier
transform techniques (Binney and Tremaine 1987, section 2.8).

The validity of the constant $M/L$ assumption can be evaluated from
detailed rotation curve studies, or from multi-color near-infrared
surface photometry. For example, \citet{fre92}
reported an analysis of a sample of more than 550 H$\alpha$ rotation
curves and $I$-band surface photometric profiles derived
by \citet{mat92},
which shows that the stellar mass distribution alone, with  $M/L$
$\sim$ constant, reproduces the observed optical rotation curves
for about 97 percent of their sample. The rotation curve morphologies
cited by Freeman span the entire range: from almost entirely solid
body to almost entirely flat. In a more recent study, \citet{per96}
corroborate this result,
and show that the optical rotation curves of spiral galaxies may be fitted by
a constant mass-to-light ratio (except for dwarf galaxies, none of
which we study here).

Constraints on the dark halo content of barred galaxies
may also be deduced from the dynamical friction or drag of bars
rotating within dark matter halos. The studies of Debattista and
Sellwood (1998, 2000) indicate that bars are only able to maintain
their high pattern speeds if the disk itself  provides
most of the gravitational potential -- a high central density, dark matter
halo would simply provide too much drag on the bar.
Such independent studies suggest that light effectively traces
mass within the optical disks of barred spiral galaxies. A detailed
dynamical study of one barred spiral, NGC 4123, by \citet{wei98} shows
that the inner regions of the disk must be at 90-100\% of the required
``maximum disk'' forcing.

The method and program that we will use to derive potentials for
our sample galaxies is described
by \citet{qui94}, hereafter QFG. These authors describe a practical
application of the convolution equations which can be used on
two-dimensional images and which explicitly accounts for
disk thickness. By assuming the $z$-dependence
of the mass density is independent of position in the plane, the potential
can be derived as the convolution of the image intensities
with a function that integrates over the vertical dependence of the density,
$\rho_z(z)$. The $z$-dependence of the density in an isothermal sheet
varies as $sech^2(z/2h_z)$, where $2h_z$ is the isothermal scale height.
\citet{vdk81a,vdk81b} showed that this representation applies
well to surface photometry of some edge-on galaxies. However,
\citet{wai90} found that an exponential dependence of
$\rho_z(z)$ provides a better description of the vertical light distribution
in IC 2531. In their analysis of NGC 4314, QFG estimated the exponential
scale height $h_z$ as (1/12)$h_R$, where $h_R$ is the radial scalelength.
QFG also tested the constant $M/L$ assumption by deriving the $J-K$ color
index across the disk of NGC 4314. Little color variation was found,
prompting them to conclude the constant $M/L$ assumption  was
realistic, especially in the bar region.

The issue of whether to use an exponential rather than an isothermal
density law in our analysis is a complex one. \citet{deg98} has presented
a new detailed study of highly inclined disk galaxies
in the $K'$ (2.1 $\mu$m) band. Since the near-infrared light is relatively
insensitive to contamination by galactic dust, he has followed the vertical light
distributions down to the galactic planes. He finds that such distributions are
more peaked than expected for a $sech(z)$ distribution, but rounder than an
exponential function. However, he has demonstrated that it is possible for all
the galaxies in his sample to have intrinsically exponential vertical surface
brightness distributions. \citet{elb99} have
demonstrated that disks which are exponential in both the radial and
perpendicular directions give excellent fits to the
profoundly asymmetric red-to-blue $V-K$ color profiles found
along the minor axes of many spiral galaxies.

The scaleheights of our largely face-on galaxies are not known, thus
we can only assume values in our analysis. As noted by \citet{fre96},
some bars are thicker than their disks, while others are probably
about the same thickness. He notes that the bar/bulge of our Galaxy
has a similar exponential scale height to the old disk. The average blue
absolute magnitude of the galaxies in our sample is $-$20.5 $\pm 0.8$,
comparable to the Milky Way \citep{dev78}.
Therefore, for our {\it preliminary} analysis,
we adopt, for all of our sample galaxies, $h_z$ = 325 pc, which is the
exponential scaleheight of our Galaxy \citep{gil83}.  Not all
galaxies do have the same exponential scaleheight: the
study by \citet{deg98} indicates that
late-type galaxies on average have a thinner disk than earlier type
systems.  To account for possible variations in scaleheight based on
morphological
type, and for the possibility that some bars are thicker than their disks,
we have made separate potential runs for $h_z$ = 225pc and
425pc.
In the future, it should be possible to improve our judgment of $h_z$
by scaling from values of the radial scalelength, as done by QFG.
However, from de Grijs's work, the ratio $h_R/h_z$ has a rather large
scatter, ranging from 4 to 12 over the type range Sb to Sd, and
from 2.5 to 7 at type Sb alone.

Many of the galaxies in our sample retain their
bar strength class values irrespective of scaleheight variations from 225pc
to 425pc. When a galaxy does move from one bar strength to the next,
the effect of decreasing the scaleheight is to increase the bar strength;
increasing the scaleheight leads to a decrease in the bar strength. These
effects were also noted by QFG in their study of NGC 4314
and by \citet{sal99} in their study of IC 4214.
In our analysis, an uncertainty of $\pm$100pc in $h_z$
produces an average uncertainty of $\mp$13\% in bar strength.

\section{Calculating the Bar Strength}

Our procedure for calculating the bar strength involved several steps:
(1) deprojection of a sky-subtracted $H$ or $K$-band image, using position
angles and isophotal axis ratios from RC3 \citep{dev91},
except for the ringed galaxies NGC 1326, 1433, 3054, 3081, 6300,
ESO 566$-$24, and ESO 565$-$11,
where the deprojections are based on deep photometric
and optical or HI kinematic studies \citep{but87,ryd96,but98,but00,pur98};
(2) expansion
or contraction of the array to dimensions of 2$^n$, as required by the
QFG fast Fourier transform analysis; (3) calculation of the potential
with an arbitrary mass-to-light ratio of 1.00, a scaleheight of 225, 325, or
425 pc, and a distance from \citet{tul88} or (for galaxies not in this
catalogue) from the linear Virgocentric flow model \citep{aar82}
and a Hubble constant
of 75 km s$^{-1}$ Mpc$^{-1}$; (4) derivation of the $m$ = 0, or axisymmetric,
part of the derived potential; (5) calculation of the mean axisymmetric
radial ($<F_{R}>$) and tangential
($F_T$) force fields; and (6) computation of the ratio map,

\begin{equation}
Q_T(i,j) = {F_T(i,j)\over <F_{R}(i,j)>}.
\end{equation}

\noindent
Note that the distance is needed only because we are assuming scaleheights
in parsecs. If we could infer $h_z$ as a specific fraction of the radial
scalelength $h_R$, then no distance would be required. Also, the analysis
is independent of the assumed (constant) value of the mass-to-light ratio
because we are computing a force ratio. The actual force fields would
scale the same way with the mass-to-light ratio.

The ratio map is our principal tool for estimating the bar strength.
Figure 1 (top left) shows a typical ratio map for a strongly-barred galaxy,
NGC 1433.
The map shows four well-defined regions where the ratio reaches a maximum
or minimum around or near the ends of the bar. This pattern is the
{\it characteristic signature} of a bar. Depending on quadrant relative
to the bar, these regions can be negative or positive, because the sign
of the tangential force changes from quadrant to quadrant (see Figure 1,
top middle).  The absolute values of the ratio at these points
represent the maxima of $Q_T$ in the \citet{com81} formulation.
To locate the maxima automatically, we analyze azimuthal profiles (5$^{\circ}$
steps) of
$Q$ as a function of radius, and then derive $Q_T$ at each radius.
Then, we derive the value and location of $Q_T^{max}$
in each quadrant. Let $Q_{bi}$ be the value of $Q_T^{max}$ in quadrant
$i$. Then we define the bar strength as

\begin{equation}
Q_b = \sum_{i=1}^4 Q_{bi}/4.
\end{equation}

\noindent
We analyze $Q_b$ by quadrant because few galaxies are so symmetric
that $Q_T$ reaches its maximum at the same radius and value in all
quadrants. For example, the presence of a strong spiral can make the values
of $Q_b$ somewhat different in the trailing quadrants of the
bar compared to the leading quadrants, or some asymmetry in the bar
itself can make the regions unequal. We use the mean error in $Q_b$
as a measure of how well the different quadrants agreed in bar
strength.

Table 1 defines the bar strength classes that we base on the measured
values of $Q_b$. Except for class 0, each class spans a range of 0.1
centered on a unit value. Thus, bar class 1 includes galaxies having $Q_b$ =
0.1$\pm$0.05, bar class 2 includes galaxies having $Q_b$ = 0.2$\pm$0.05, etc.
With these definitions, class 0 involves a narrower range of $Q_b$,
since $Q_b$ cannot be negative as defined.

Table 2 lists the bar strengths so derived for a small sample of
representative galaxies covering spiral and ring morphologies,
and over a range of Hubble types. The parameter listed as the ``family''
in Table 2 is from \citet{dev63} for all galaxies in the table except
IC 4290, ESO 565$-$11, ESO 566$-$24, and Maffei 2. The optical
bar classifications
for IC 4290 and ESO 566$-$24 are from \citet{but98}, that for ESO 565$-$11
is from \citet{bpc95}, and that for Maffei 2
is from \citet{bum99}.  Using the code in Table 1, each
galaxy in Table 2 was assigned into its appropriate bar strength class.
The strongest bars we find reach bar class 6, where the maximum tangential
force reaches about 60\% of the mean radial force.
Figure 2 shows how well $Q_b$ correlates with the de Vaucouleurs
family parameter in the sample of Table 2. This shows that the threshold
for calling a galaxy ``SB'' seems to be bar class 2. Apparently,
bars become obvious at this strength and the Hubble-de Vaucouleurs
classification can make no further discrimination on bar strength
beyond this threshold, so that ``SB'' also includes galaxies up to
bar class 6. Galaxies classified as
``SA'' or ``SAB'' mostly range from classes 0 to 2, with the exception of
Maffei 2 which is bar class 3. Maffei 2, which is a heavily reddened
galaxy in the Zone of Avoidance, is probably the
nearest massive SBbc galaxy \citep{spi73,hur93,bum99}.

Some of the more weakly-barred galaxies in Table 2
required special treatment. For example,
the measured values of $Q_b$ for NGC 2997 and NGC 7083 indicate
a bar class of 1, but we have
specified their classes as (0) since there is either no clear bar signature
in the ratio map, or the program measured obvious
deprojection stretch of the bulge.

The uncertainty attached to each
value of $Q_b$ in Table 2 includes the mean error in $Q_b$ from the
four quadrants and the uncertainty of $\pm$100pc due to scale height.
It excludes deprojection uncertainties, however. We consider the $Q_b$
values in Table 2 preliminary, because orientation parameters
of many of the galaxies are manifestly improvable. For our analysis,
we mostly used deprojected images that were already available to us,
many of which were previously used for defining the dust-penetrated pitch
angle classes described by \citet{blp99} and \citet{blo99}.
We used well-defined (kinematically and/or photometrically-based)
orientation parameters mainly for the ringed galaxies at this time
since the deprojected images were also already available to us from earlier
studies. We also used well-defined orientation parameters for Maffei 2
from \citet{bur96} and \citet{bum99}.
Table 3 shows the impact of large uncertainties in the
orientation parameters on $Q_b$ for a representative barred galaxy, NGC 1300.
This is a special case because
the bar major axis of this galaxy is nearly along the line of nodes,
making its strength very sensitive to the assumed inclination.
Uncertainties of $\pm$10$^{\circ}$ on the inclination $i$ and line of nodes
position angle $\phi_n$ changes the bar strength over the bar class range 3-5.
The RC3 orientation parameters of $i$=49$^{\circ}$ and $\phi_n$ = 106$^{\circ}$
may be compared to the values of 50$^{\circ}$ and 95$^{\circ}$, respectively,
determined from HI kinematics by \citet{eng89}. According to Table 3,
NGC 1300 is still bar class 4 with England's parameters. In general,
whenever the bar minor axis or major axis is nearly along the line of
nodes, the effect of inclination will be most serious on $Q_b$.

Figure 1 (bottom three rows) shows nine galaxies in our sample in a
sequence of increasing $Q_b$, with the locations of the maximum points
indicated. The mean position angle difference between the maximum
points and the bar depends on the
importance of higher order terms in the bar potential. For example,
if the bar is a pure $m$ = 2 potential, these points
would lie at $\pm$45$^{\circ}$ to the bar axis. However, in most
cases, the angle we find is less than $\pm$45$^{\circ}$, because of
the importance of $m$ = 4 and 6 terms in most real bars.

\section{Evaluation of Technique}

Equation 1 offers a straightforward way of deriving
bar strength for large numbers of galaxies in an efficient, automatic
manner, subject of course to our assumptions.
Studies of individual galaxies, such as NGC 7479 \citep{sem95a},
NGC 4321 \citep{sem95b}, NGC 4314 (QFG), NGC 1300 \citep{lik96},
NGC 1365 \citep{lin96}, NGC 4123 \citep{wei98},
and NGC 1433, 3081, and 6300 \citep{bco00} {\it corroborate the M/L
  $\sim$ constant assumption}, especially in the inner regions of galaxies.
Equation 1's  principal advantage, as we see it, is that we do not need
to define the bar in any way other than to limit the analysis to
the inner regions of a galaxy to avoid strong
spiral structure or corner effects. It can be sensitive to
deprojection uncertainties for objects like NGC 1300 (Table 3)
and also in cases with roughly spherical bulges
and significant inclinations, because then the artificial stretching
of the bulge can produce a bar-like signature. However, these
are easy to identify since the bulge is stretched along the minor
axis, and these are also easy to treat with bulge decomposition
if there is little or no actual bar.

Excessive star formation could of course impact the computed
potentials. It is well-known that young red supergiants, while
on the whole contributing only 3\% of the total 2.2$\mu$m flux,
can be locally dominant in star-forming regions and contribute
up to 33\% of the flux \citep{rho98}. We tested the impact of
star formation in M100 by removing the obvious star-forming regions
in the arms and bar in a $K$-band image, assuming that their
mass-to-light ratios are much less than that of the old disk.
The removal of these objects had little impact on the estimated
bar strength, and the bar class was unchanged.

$Q_b$ does not measure the shape of an {\it isolated} bar. It also
accounts for the disk in which the bar is embedded. A comparison between
our bar classes and the bar ellipticity classes of \citet{mar95} shows
that ellipticity class 7 (one of the strongest classes in Martin's sample)
can have a bar strength class from 2 (e.g., M83) to 6 (e.g., NGC 7479),
while an ellipticity class 4 galaxy can be bar strength class 0 (e.g.,
NGC 4653). Another example of an optically  strong ellipticity bar
class 
7 galaxy 
but with a non-discernable gravitational potential
is NGC 4395 (Knapen, private communication).  

$Q_b$ correlates better with the relative
bar fraction, $L(bar)/L(R<R_{25})$, of \citet{elm85}, although
we have only three galaxies in common.
NGC 1300, 3992, and 7479 have relative bar luminosities of 9.2\%, 3.7\%, and
20\%, respectively. These same galaxies (for our adopted orientation
parameters) have $Q_b$ = 0.42, 0.35, and 0.63, respectively.
Both the Martin and Elmegreens' results are based on $B$ or $I$ images.
Our sample
does not overlap with the sample of Seigar and James (1998), and we cannot
make a comparison between $Q_b$ and bar ``equivalent angle'' at this time.

On the other hand, as we have noted, $Q_b$ can be (but is not
always) sensitive to the
scale height, {\it values of which are assumed in this prelimary
study}. we find that bar class is significantly sensitive to $h_z$
only for the strongest bars; however, even in those cases the uncertainty
will usually be less than 1 bar class.

\section{Impact of Bulge Shape}

Our tabulated values of $Q_b$ have ignored the shape
of the bulge. In each case, the bulge has been assumed to be as flat as
the disk, and only in a few cases have we treated the bulge as less flattened
than the disk in order to eliminate deprojection stretch. The assumption
of a highly flattened bulge is probably valid for some galaxies, such
as for barred spirals with triaxial bulges \citep{kor82,kor93}.
However, in many galaxies the bulge is a much less flattened component,
and even for a face-on galaxy, this could impact our bar strength
estimates. If the light distribution of a spherical bulge is transformed
into a potential assuming it is a thin disk, then the axisymmetric
radial forces derived will be too large, especially in the inner regions.
To evaluate this, we modeled the bulge of one of our galaxies, NGC 1433,
whose bulge can be interpreted in terms of a highly flattened triaxial
inner section and a more spherical outer section \citep{but00}.
The apparently round part of the
bulge of NGC 1433 includes 27\% of the total $H$-band luminosity
and can be modeled as
a double exponential with scale lengths of 2\arcs\ and 11\arcs, both
much less than the 80\arcs\ radius of the bar. Using a program that
derives forces from the light distribution by
modeling the density in terms of spheroids of different flattenings
\citep{kal86}, we find that our
analysis overestimates the radial forces in NGC 1433 mainly for radii
less than 5\arcs\ where the error can reach a factor of two. At the
radius of the bar, the effect of the bulge shape is likely to be
negligible. This may not be true in every case, and we will consider
the influence of bulge shape in individual cases in more detail in a
separate study.
However, we note that it would be incorrect to measure bar strength
only on bulge-subtracted images, because the bulge contributes to the
axisymmetric background regardless of its shape, and $Q_b$ should measure
the strength of the bar relative to all axisymmetric luminous components.

\section{A Dust-Penetrated Quantitative Classification of Spiral
  Galaxies}

The Hubble classification of galaxies is based on their optical
appearance. As one moves from early to late type spirals, both unbarred
and barred, the appearance will be dominated more and more by the young
Population I component of gas and dust. However, the gaseous Population
I component may only constitute 5 percent of the dynamical mass of a
galaxy \citep{fro96}.

The Hubble classification does not
constrain the {\it dynamical mass distribution}, as corroborated by 
\citet{bur85}, \citet{blo96} and \citet{blo99}.
Galaxies with bulge-disk ratios differing
by a factor of 40 can have very similar shapes to their rotation
curves; galaxies of Hubble type a, for example, may
belong to any one of three different mass classes \citep{bur85}.
A late type galaxy such as NGC 309 (optical type c) may co-exist
with an early type a evolved disk morphology \citep{blow91}.
The decoupling
between gaseous and stellar disks can be profound, as reviewed
by \citet{blo99}.

A classification scheme of spiral galaxies in the near-infrared
was recently proposed by \citet{blp99}. The main classification
parameter is the dominant Fourier harmonic in the
spiral arms. In this classification, a ubiquity of
low order ($m$=1,2) Fourier modes for both normal and barred galaxies is
found in the near-infrared regime, consistent with the modal theory
of spiral structure \citep{ber96,blo99}. Galaxies with
a dominant Fourier m=1 mode are L=lopsided, while galaxies principally
showing an m=2 harmonic are E=evensided. 
For simplicity ofcourse, any harmonic class can simply be denoted by
H$m$, where lopsided and evensided galaxies carry the H1 and H2
designations respectively.
\citet{blo99} bring attention those rarer galaxies
with higher order dominant harmonics (e.g., NGC 5054, $m$=3, and ESO 566$-$24,
$m$=4) which are assigned to harmonic classes H3 and H4,
respectively.

Within these dust-penetrated harmonic classes, galaxies are binned into
three subclasses based on the pitch angle of the spiral arms, robustly
determined from Fourier spectra. Evolved stellar disks with
tightly wound spiral arms characterized by near-infrared pitch angles
of $\sim$10$^{\circ}$ are binned into the $\alpha$ class; the $\beta$
class is an intermediate group, with near-infrared pitch angles of
$\sim$25$^{\circ}$, while open stellar arms with pitch angles of
$\sim$40$^{\circ}$ define the $\gamma$ class.

These
$\alpha$, $\beta$ and $\gamma$ classes are inextricably related to the
rate of shear A/$\omega$ in the stellar disk. Here A is the first Oort
constant and $\omega$ is the angular velocity. Falling rotation curves
generally give rise to the $\alpha$ class, while rising rotation
curves give rise to the $\gamma$ class. For a complete
discussion, see \citet{blo99}. 
 When
imaged in the near-infrared, a Hubble or de Vaucouleurs early type b
galaxy may either belong to class $\alpha$ or $\beta$ or $\gamma$.
Likewise for the other optical Hubble bins. Hubble type and dynamical
mass distributions are not correlated \citep{bur85}.

In this paper, we propose to simply add the bar strength
derived from the inferred gravitational force fields to the shear-related
pitch angle classes of \citet{blp99}, in order
to define a more complete dust-penetrated classification system. Class 0
spirals have no bar or oval; our strongest bars are class 6.
Figure~\ref{tuning} shows a quantitative
three-pronged ``tuning fork'' for the near-infrared images of nine spiral
galaxies that illustrate the full dust-penetrated classification
system for the H2 class.
The images are from \citet{blo99} and are overlaid with the contours of
the main harmonic, $m$=2.

For example, it is proposed that NGC 5236=M83 (illustrated in Figure
3) bear the quantitative classification H2$\alpha$2, implying that the
bar class is 2, the galaxy has tightly wound type $\alpha$ arms in the
dust-penetrated regime, and that there is an evensided two-armed
spiral in its evolved stellar disk. The pitch angles of the
spiral arms in barred galaxies of the same Hubble type span the entire $\alpha$ to $\gamma$ range:
for example, NGC 3992 and NGC 1365 (both Hubble type b)
carry the full designation  H2$\alpha$4 and H2$\gamma$5 respectively.

The tuning fork in Figure 3 may serve as a z$\sim$0 template when
galaxies at z $\sim$ 0.5-1 are imaged in their restframe $K$ dust
penetrated 2$\mu m$ regime. It is not a confirmed observational fact
that the morphology of galaxies in the Hubble Deep Fields are very
different in the past than in the present (R.I. Thompson, private
communication) when the effects of redshift
and surface brightness dimming are fully accounted for 
\citep{ellis97,tak99}.

\section{Conclusions}

We have outlined the derivation of
a fully quantitative bar strength class for spiral galaxies, based on
the the maximum value of the amplitude of the tangential force to the
mean radial force. Although
the method is still based on light fluxes, since we use images to infer
gravitational potentials, it provides a more direct handle on bar strength
than any other light-based methods so far applied. Regardless of what the bar
may visually look like, the ratio map will show a pattern of four maxima/minima
that can be isolated fairly automatically to give a robust measure of bar
strength. The method is thus free from uncertainties
connected with defining bars or of full bulge/disk
decompositions. It can be applied quickly and efficiently to many
galaxies, and only in some cases is it necessary to give special
treatment to the bulge if it suffers too much deprojection stretch.
Bulge shape, dark matter, and star formation
have little impact on the bar strength class for our sample galaxies.
The most important effects are vertical scale height and,
for highly inclined galaxies, deprojection uncertainties.

The method has much room for refinement. For example, bulges
can be decomposed to eliminate deprojection stretch and their influence
on the ratio maps more accurately accounted for. Improved orientation
parameters can be used or derived for galaxies as data become available.
Appropriate vertical scaleheights could be inferred more reliably from detailed
near-infrared surface photometry and the type dependence of $h_R/h_z$.
The validity of the constant $M/L$
assumption could be tested by measuring $J-K$ color distributions.
Finally, deprojection uncertainties
go beyond simple bulge deprojection stretch to the fundamental uncertainty
of deprojecting any galaxy, even if a bulge is not present. All of these
are issues which will be considered in a more detailed forthcoming study.

The addition of our bar strength parameter to the dust-penetrated
classification scheme of \citet{blo99} now gives a full
dynamical appreciation of the range of evolved stellar disks with bars
(Figure~\ref{tuning}).

We thank A. Quillen, F.Combes, A. Kalnajs, and S. M. G. Hughes
for programs that were used in this study. We also thank A. Quillen,
F.Combes, B.Elmegreen, K. Freeman, P. van der Kruit, A. Kalnajs, R. de Grijs,
R.I. Thompson and especially the (anonymous) referee for helpful and timely
advice and comments which improved this study. D. Elmegreen, I.
Puerari, P.Grosb{\o}l, J. Knapen,
K. C. Freeman, O. K. Park, and T. Jarrett very graciously loaned us some
images which we used to test the bar strength method. We thank
D. A. Crocker for her assistance in producing the ``tuning fork'' diagram
in Figure~\ref{tuning}. We gratefully acknowledge the support of the Anglo-American
Chairman's Fund, which also made the visit to South Africa of one of us
(RB) possible. We are indebted to CEO Mrs. M. Keeton and the Board of Trustees.

\begin{table}
\caption{Bar Strength Classes}
\begin{tabular}{cl}
\tableline\tableline
Class & Range in $Q_b$ \\
\tableline
0  & $<$0.05 \\
1  & 0.05--0.149 \\
2  & 0.15--0.249 \\
3  & 0.25--0.349 \\
4  & 0.35--0.449 \\
5  & 0.45--0.549 \\
6  & 0.55--0.649 \\
\tableline
\end{tabular}
\end{table}

\begin{table}
\caption{Bar Strengths for 36 Galaxies}
\begin{tabular}{llcccllccc}
\tableline\tableline
Galaxy & Family\tablenotemark{a} & $Q_b$ & Bar & DP & Galaxy & Family\tablenotemark{a} & $Q_b$ & Bar & DP \\
       &        &       & Class & Type & &        &       & Class & Type \\
\tableline
N0309     & SAB &    .11$\pm$   .02 &      1  & H2$\beta$1 & N4548    &
SB  &    .44$\pm$   .03 &      4  & H2$\alpha$4 \\
N0521     & SB  &    .18$\pm$   .03 &      2  & H2$\alpha$:2 & N4622    & SA  &    .01$\pm$   .01 &      0  & H1$\alpha$0 \\
N0718     & SAB &    .15$\pm$   .02 &      2  & H2$\beta$2 & N4653    & S$\underline {\rm A}$B &   .04$\pm$   .01 &      0 & H2$\beta$0 \\
N1300    & SB  &    .42$\pm$   .06 &      4  & H2$\alpha$4 & N4902
& SB  &    .29$\pm$   .04 &      3  & H2$\alpha$3 \\
N1326    & SB  &    .16$\pm$   .02 &      2  & H2$\alpha$2 & N5236    & SAB &    .19$\pm$   .04 &      2  & H2$\alpha$2 \\
N1365    & SB  &    .46$\pm$   .07 &      5  & H2$\gamma$5 & N5371    & S$\underline {\rm A}$B &    .19$\pm$   .02 &      2 & H2$\gamma$2 \\
N1433    & SB  &    .38$\pm$   .05 &      4  & H2$\alpha$4 & N5457    & SAB &    .12$\pm$   .01 &      1  & H2$\alpha$1 \\
N1637    & SAB &    .09$\pm$   .03 &      1  & H1$\gamma$1 & N5905    & SB  &    .43$\pm$   .05 &      4  & H2$\gamma$4 \\
N2543    & SB  &    .28$\pm$   .05 &      3  & H2$\beta$3 & N5921    & SB  &    .38$\pm$   .04 &      4  & H2$\gamma$4 \\
N2857    & SA  &    .09$\pm$   .02 &      1  & H2$\alpha$1 & N6300    & SB  &    .17$\pm$   .03 &      2  & H2$\beta$2  \\
N2997\tablenotemark{b}    & SAB &    .06$\pm$   .02 &      (0) & H2$\beta$0 & N6782    & SAB &    .19$\pm$   .02 &      2  & H2$\alpha$2 \\
N3054    & SAB &    .17$\pm$   .02 &      2  & H2$\alpha$2 &
N7083\tablenotemark{c}    & SA &    .06$\pm$   .01 &     (0)  & H2$\gamma$0 \\
N3081    & SAB &    .17$\pm$   .02 &      2  & H2$\alpha$2 & N7098    & SAB &    .20$\pm$   .02 &      2  & H2$\alpha$2 \\
N3346    & SB  &    .25$\pm$   .06 &      3  & H2$\beta$3 & N7479    & SB  &    .63$\pm$   .08 &      6  & H2$\gamma$6 \\
N3992    & SB  &    .35$\pm$   .05 &      4  & H2$\alpha$4 & I4290     & SB  &    .56$\pm$   .08 &      6  & H2$\alpha$6 \\
N4051    & S$\underline {\rm A}$B &  .24$\pm$   .05 &      2 & H2$\gamma$2 & E565-11  & SB  &    .28$\pm$   .03 &      3  & H2$\beta$3 \\
N4321    & SAB &    .12$\pm$   .03 &      1  & H2$\beta$1 & E566-24  & SB  &    .27$\pm$   .04 &      3  & H4$\beta$3 \\
N4394    & SB  &    .22$\pm$   .04 &      2  & H2$\alpha$2 & Maffei 2 & SAB &   .27$\pm$   .04 &      3  & H2$\gamma$3 \\
\tableline
\end{tabular}
\tablenotetext{a}{From \citet{dev63}, except for Maffei 2, IC 4290,
ESO 565$-$11, and ESO 566$-$24 (see text)}
\tablenotetext{b}{No clear bar signature in ratio map;
correct bar class in parentheses}
\tablenotetext{c}{Measured amplitude due to deprojection stretch;
correct bar class in parentheses}
\end{table}

\begin{table}
\caption{Effect of Uncertainty in Orientation
Parameters on $Q_b$ of NGC1300}
\begin{tabular}{cccc}
\tableline\tableline
$i$/$\phi_n$ & 96$^{\circ}$ & 106$^{\circ}$ & 116$^{\circ}$ \\
\tableline
39$^{\circ}$  &    0.51  &    0.50  &   0.52 \\
49$^{\circ}$  &    0.44  &    0.42  &   0.45 \\
59$^{\circ}$  &    0.32  &    0.30  &   0.35 \\
\tableline
\end{tabular}
\end{table}

\figcaption[]{Illustration of technique: (row 1, left) Ratio map for
NGC 1433 ($H$-band), showing maximum points in $Q_b$ (filled white
squares); (row 1, middle) Schematic of radial and tangential forces in NGC 1433
at the maximum points (axes arcseconds for bar schematic only); (row 1, right) image of NGC 1433 with maximum points
superposed (scales of these panels are all slightly different); Rows
2--4, a sequence of galaxies of increasing $Q_b$.
(row 2, left to right): NGC 4622 (0.01), NGC 309 (0.11), NGC 3081 (0.17);
(row 3, left to right): M83 (0.19), NGC 4902 (0.29), NGC 3992 (0.35);
(row 4, left to right): NGC 1365 (0.46), IC 4290 (0.56), NGC 7479 (0.63).
The filled squares again show the locations of the maximum points.
\label{fig1}}

\figcaption[]{Plot of $Q_b$ versus de Vaucouleurs family parameter for
Table 2 galaxies,
where the family is taken mainly from \citet{dev63} (see text for other
sources).
\label{fig2}}

\figcaption[]{A quantitative fork for z$\sim$0 spiral
galaxies in their near-infrared dust-penetrated regime. Galaxies are
binned according to three quantitative criteria: H$m$, where $m$ is the
dominant Fourier harmonic (illustrated here are the two-armed H2
family); the pitch angle families $\alpha$, $\beta$ or $\gamma$
determined from the Fourier spectra, and thirdly the
bar strength, derived from the gravitational torque (not ellipticity)
of the bar. Early type b spirals (NGC 3992, NGC 2543, NGC 7083, NGC
5371 and NGC 1365) are distributed within all three families
($\alpha$, $\beta$ and $\gamma$). Hubble type and dust-penetrated class
are uncorrelated.
\label{tuning}}

\end{document}